\documentclass{article}

\usepackage{PRIMEarxiv}
\usepackage{float} 

\usepackage[utf8]{inputenc} 
\usepackage[T1]{fontenc}    
\usepackage{hyperref}       
\usepackage{url}            
\usepackage{booktabs}       
\usepackage{amsfonts}       
\usepackage{nicefrac}       
\usepackage{microtype}      
\usepackage{lipsum}
\usepackage{graphicx}       
\usepackage{amsmath}       
\usepackage{listings}
\usepackage{xcolor}
\usepackage{fancyhdr}       
\usepackage{graphicx}       
\graphicspath{{media/}}     
\usepackage{caption}
\usepackage{multirow}
\usepackage{soul}
\usepackage{enumitem}
\setlist[itemize]{leftmargin=*,itemsep=0pt, topsep=0pt, parsep=0pt}
\usepackage{pdfpages}

\usepackage[most]{tcolorbox} 
\usepackage{fancyvrb}        
\usepackage{xcolor}          
\usepackage{soul}            

\usepackage{makecell}
\usepackage{tabularx}

\definecolor{aigold}{RGB}{244,210, 1} 
\definecolor{aigreen}{RGB}{210,244,211} 

\sethlcolor{aigreen}

\definecolor{aired}{RGB}{255,180,181}

\definecolor{aigold}{RGB}{255,180,181}

\definecolor{aiblue}{RGB}{173,216,230} 
 
\definecolor{lightred}{rgb}{1,0.9,0.9} 

\newcounter{textbox}
\tcbset{
  custombox/.style={
    width=512.18663pt,
    top=3pt,
    fonttitle=\bfseries\small\sffamily, 
    colback=white,
    colframe=black,
    colbacktitle=black,
    boxrule=0.5pt, 
    colback=white, 
    enhanced,
    rounded corners, 
    center,
    boxed title style={
            sharp corners,
            size=small,
            colframe=blue!75!black, 
        },
    attach boxed title to top left={yshift=-0.1in,xshift=0.15in},
    boxed title style={boxrule=0pt,colframe=white,},
    before=\setlength{\baselineskip}{-2pt}, 
    before upper=\setlength{\baselineskip}{8pt}, 
    after=\setlength{\baselineskip}{8pt}, 
  }
}
\newtcolorbox{LLMbox}[2][]{custombox, width=\textwidth, title=#2, #1}

\newtcbox{\mybox}[1][green]{on line,
arc=0pt,outer arc=0pt,colback=#1!10!white,colframe=#1!50!black,
boxsep=0pt,left=0pt,right=0pt,top=0pt,bottom=0pt,
boxrule=0pt,bottomrule=0pt,toprule=0pt}

\tcbset{
  customboxmultipage/.style={
    width=\textwidth, 
    top=2pt, 
    bottom=2pt, 
    left=2pt, 
    right=2pt, 
    fonttitle=\bfseries\small\sffamily, 
    colback=white, 
    colframe=black, 
    boxrule=0.5pt, 
    breakable, 
    rounded corners, 
    center, 
    boxed title style={
      sharp corners,
      size=small,
      colframe=blue!75!black, 
      colback=white, 
    },
    attach boxed title to top left={
      yshift=-0.1in, 
      xshift=0.15in  
    },
    boxed title style={
      boxrule=0pt, 
      colframe=white, 
    },
    enhanced, 
    nobeforeafter, 
  }
}

\newtcolorbox{LLMboxmultipage}[2][]{customboxmultipage,title=#2,#1}

\lstdefinelanguage{json}{
    basicstyle=\ttfamily\scriptsize, 
    numbers=left,
    numberstyle=\tiny\color{gray}, 
    stepnumber=1,
    numbersep=10pt, 
    showstringspaces=false,
    breaklines=true,
    frame=none,
    backgroundcolor=\color{white},
    literate=
     *{0}{{{\color{blue}0}}}{1}
      {1}{{{\color{blue}1}}}{1}
      {2}{{{\color{blue}2}}}{1}
      {3}{{{\color{blue}3}}}{1}
      {4}{{{\color{blue}4}}}{1}
      {5}{{{\color{blue}5}}}{1}
      {6}{{{\color{blue}6}}}{1}
      {7}{{{\color{blue}7}}}{1}
      {8}{{{\color{blue}8}}}{1}
      {9}{{{\color{blue}9}}}{1}
      {:}{{{\color{red}:}}}{1}
      {,}{{{\color{red},}}}{1}
      {\{}{{{\color{brown}\{}}}{1}
      {\}}{{{\color{brown}\}}}}{1}
      {[}{{{\color{brown}[}}}{1}
      {]}{{{\color{brown}]}}}{1},
}

\lstdefinelanguage{yaml}{
  basicstyle=\ttfamily\footnotesize, 
  keywordstyle=\color{blue},
  comment=[l]{\#},
  commentstyle=\color{gray},
  morestring=[b]',
  morestring=[b]",
  stringstyle=\color{orange},
  sensitive=true,
}

\title{
\bf{Bioinspired123D: Generative 3D Modeling System for Bioinspired Structures}
\thanks{\textit{\underline{Citation}}: 
\textbf{R.K. Luu, et al., Title. Pages.... DOI:000000/11111.}} 
}

\author{
    Rachel K. Luu \\
  Department of Materials Science and Engineering \\
   Laboratory for Atomistic and Molecular Mechanics (LAMM) \\
  Massachusetts Institute of Technology\\
  Cambridge, MA, USA\\
  https://orcid.org/0000-0002-7821-934X
   \And
  Markus J. Buehler \\
  Department of Civil and Environmental Engineering\\
  Department of Mechanical Engineering \\  
  Center for Computational Science and Engineering \\
  Schwarzman College of Computing \\  
  Laboratory for Atomistic and Molecular Mechanics (LAMM) \\
  Massachusetts Institute of Technology \\
  Cambridge, MA, USA\\ 
  https://orcid.org/0000-0002-4173-9659 \\
  \\
  Corresponding author: \texttt{mbuehler@MIT.EDU} 
}

\begin{document}
\maketitle

\begin{abstract}
Generative AI has made rapid progress in text, image, and video synthesis, yet text-to-3D modeling for scientific design remains particularly challenging due to limited controllability and high computational cost. Most existing 3D generative methods rely on meshes, voxels, or point clouds which can be costly to train and difficult to control. We introduce Bioinspired123D, a lightweight and modular code-as-geometry pipeline that generates fabricable 3D structures directly through parametric programs rather than dense visual representations. At the core of Bioinspired123D is Bioinspired3D, a compact language model finetuned to translate natural language design cues into Blender Python scripts encoding smooth, biologically inspired geometries. We curate a domain-specific dataset of over 4,000 bioinspired and geometric design scripts spanning helical, cellular, and tubular motifs with parametric variability. The dataset is expanded and validated through an automated LLM-driven, Blender-based quality control pipeline. Bioinspired3D is then embedded in a graph-based agentic framework that integrates multimodal retrieval-augmented generation and a vision–language model critic to iteratively evaluate, critique, and repair generated scripts. We evaluate performance on a new benchmark for 3D geometry script generation and show that Bioinspired123D demonstrates a near fourfold improvement over its unfinetuned base model, while also outperforming substantially larger state-of-the-art language models despite using far fewer parameters and compute. By prioritizing code-as-geometry representations, Bioinspired123D enables compute-efficient, controllable, and interpretable text-to-3D generation, lowering barriers to AI driven scientific discovery in materials and structural design.
\end{abstract}


\section{Introduction}
Controllable 3D structural design remains a fundamental challenge for generative AI, particularly in scientific settings where geometry, internal structure, and function are tightly coupled. This challenge is especially pronounced in biological materials when it comes to the creation of bioinspired designs.

Biological materials offer rich examples of hierarchical and functional structural design, including helicoidal plies, gradient cellular networks, and tubular architectures that tightly couple geometry with mechanical performance\cite{Buehler2013Materiomics:Structures,Meyers2008BiologicalProperties, Meyers2013StructuralConnections, Eder2018BiologicalDiversity}. Designing synthetic analogs of these systems requires navigating high dimensional shape spaces \cite{Wegst2015BioinspiredMaterials} that are generally time-consuming to explore manually. Our previous work introduced BioinspiredLLM \cite{Luu2023BioinspiredLLM:Materials}, a domain finetuned large language model (LLM) capable of generating textual descriptions of new bioinspired material concepts. In prior work, we showed that these 1D descriptions can be passed to text-to-image based diffusion models to produce 2D visual concepts. However, extending language driven design from 1D text to full 3D geometry requires methods that can translate semantic intent into spatial structure while remaining lightweight, modular, and compatible with limited compute resources.

Bridging this gap requires a different approach from traditional 3D generative models. Early work in 3D generation largely focused on image to shape pipelines based on convolutional neural networks and generative adversarial networks \cite{Han2021Image-basedEra,Chan2022EfficientNetworks,Or-El2022StyleSDF:Generation}. With the rise of modern generative AI, diffusion models emerged as a powerful framework for 3D generation, spanning voxel, point cloud, and mesh based representations \cite{Luo2021DiffusionGeneration,Zeng2022LION:Generation,Nichol2022Point-E:Prompts,Huang2025SPAR3D:Images}. More recently, image conditioned and multiview diffusion models have demonstrated impressive performance by generating multiple views from a single image \cite{Shi2023Zero123++:Model}, which are subsequently assembled into 3D meshes using dedicated reconstruction networks \cite{Xu2024InstantMesh:Models,Xiang2024StructuredGeneration}. While image based and multiview diffusion pipelines have achieved strong performance in 3D object generation, they typically require multi-stage inference and substantial computational resources. For scientific design tasks that prioritize internal structure, parametric control, and lightweight deployment over photorealism, these compute intensive pipelines are poorly matched.

As an alternative, text driven 3D generation offers a more direct and computationally efficient pathway for translating high level design intent into geometry, especially when bioinspired structures are often described through natural language. Recent work has explored leveraging LLMs to directly generate 3D meshes \cite{Wang2024LLaMA-Mesh:Models}. In parallel, code based generation approaches translate text prompts into executable scripts, enabling explicit geometric control and interpretability. Systems such as BlenderLLM focus on generating generic Blender objects and scenes \cite{Du2024BlenderLLM:Self-improvement}, while computer-aided design oriented approaches such as CAD-Coder target the generation of engineering components using Python script \cite{Doris2025CAD-Coder:Generation}. Building on these successes, an open question is how similar text to script generation strategies can be adapted to bioinspired materials, where structures are hierarchical, multiscale, and often specified through descriptive biological language rather than formal design constraints.

\begin{figure}[h]
    \centering
    \includegraphics[width=1\linewidth]{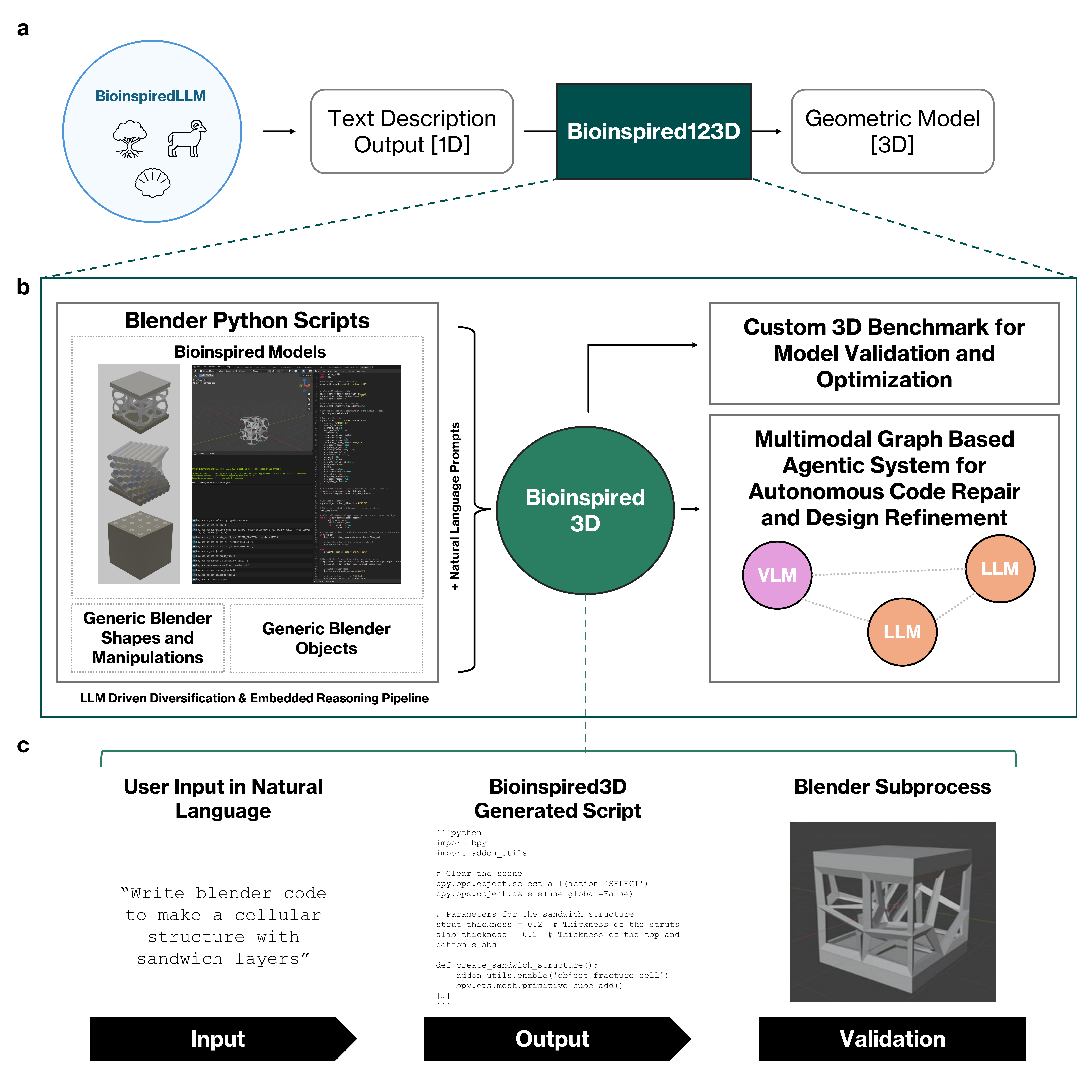}
    \caption{Study overview. a) End to end translation from 1D text descriptions to 3D geometric structures by coupling BioinspiredLLM with the Bioinspired123D pipeline. b) Overview of the Bioinspired123D system, centered on Bioinspired3D, a language model finetuned on a curated dataset of Blender Python scripts and natural language prompts, evaluated using a custom 3D benchmark and integrated within a multimodal, graph based agentic framework. c) Detailed view of Bioinspired3D, highlighting its input representation as natural language prompts and its output representation as executable Blender Python scripts. Generated scripts are extracted and validated by execution within a Blender subprocess.}
    \label{fig:Fig_1}
\end{figure}

In this work, we introduce Bioinspired123D, a compact pipeline that converts natural language prompts into 3D geometries based on bioinspired structural motifs. Rather than generating 3D meshes or point clouds directly, our approach leverages Blender’s Python interface to express geometry through programmatic instructions. This representation reduces token overhead, leverages the prior coding knowledge already embedded in foundational LLMs\cite{Wang2023AEvaluation}, and enables smooth, continuous structures well suited for bioinspired designs. With the long term goal of integrating language based materials design knowledge with geometric generation, Bioinspired123D is designed to interface seamlessly with BioinspiredLLM, enabling a pipeline from textual material descriptions to executable 3D geometry, as illustrated in Figure \ref{fig:Fig_1}a.

Bioinspired123D, shown schematically in Figure \ref{fig:Fig_1}b, consists of a new dataset, a finetuned LLM, an evaluation benchmark, and an agentic refinement framework. To support finetuning, we construct a first of its kind dataset of bioinspired 3D structures. We further augment this dataset with general purpose Blender manipulation examples and the BlendNet \cite{Du2024BlenderLLM:Self-improvement} collection to strengthen geometric reasoning. To improve robustness and diversity, the dataset is expanded using an LLM driven pipeline that also embeds step by step narrative reasoning within the generation process. A headless Blender validation pipeline ensures that all dataset entries execute successfully and produce valid rendered geometry. We then finetune a compact language model on this dataset, referred to as Bioinspired3D. Given a natural language description of a bioinspired design, Bioinspired3D outputs a Blender Python script that can be automatically validated and rendered in a Blender subprocess, as shown in Figure \ref{fig:Fig_1}c. We evaluate Bioinspired3D using a new benchmark designed to assess 3D reasoning, parameter control, and Blender script generation quality. Bioinspired3D is further integrated into an agentic system\cite{Ghafarollahi2024SciAgents:Reasoning,Buehler2024GenerativeDesign,Buehler2025TowardsValidation} that incorporates multimodal retrieval and a vision language model to critique and iteratively refine generated scripts, improving generation stability, fidelity, and geometric correctness.

Finally, we demonstrate a full text to 3D pipeline by integrating BioinspiredLLM with Bioinspired123D. In this setting, a user provides a plain text description of a biological material, such as bamboo, crab shell, or horse hoof. BioinspiredLLM then reasons over the biological structure to generate a design prompt, which is then passed to Bioinspired123D to produce a corresponding 3D bioinspired structure. Together, these results show that controllable bioinspired 3D generation can be achieved without large scale 3D foundation models. This work presents a lightweight, modular, and extensible framework for translating natural language design intent into physical, fabricable 3D structures suitable for scientific exploration, materials design, and downstream generative workflows.

\section{Results and Discussion}
\subsection{3D Bioinspired Dataset}
\label{sec:Dataset}
The design of bioinspired 3D structures is traditionally a methodical, small scale operation in studies where biological structures are first interpreted and then reproduced oftentimes using computer aided design software focusing on a few parameters of interest to understand the mechanical behavior of such design motifs \cite{Lazarus2022EquineDesigns,Lazarus2023Jackfruit:Resistance, Tee2021FromQuill}. These structures are commonly used for computational simulations\cite{Nepal2022HierarchicallyNanocomposites, Maghsoudi-Ganjeh2021ComputationalComposites, Yang2025ArtificialMechanics} or, when feasible, fabricated via 3D printing for experimental evaluation\cite{Gu2016Three-dimensional-printingComposites,He2024ReviewApplications, Shen2022Nature-inspiredLearning}. The dataset introduced in this work was generated using domain knowledge from the field of biological and bioinspired materials. Prior studies have identified eight common biological design elements that recur across natural structural systems \cite{Naleway2015StructuralBioinspiration}. In this work, we focus on three representative classes of biological architectures: helical, cellular, and tubular. Within each class, the dataset includes structured parametric variations that incorporate additional design motifs such as fibrous and layered features.

\begin{figure}[h]
    \centering
    \includegraphics[width=1\linewidth]{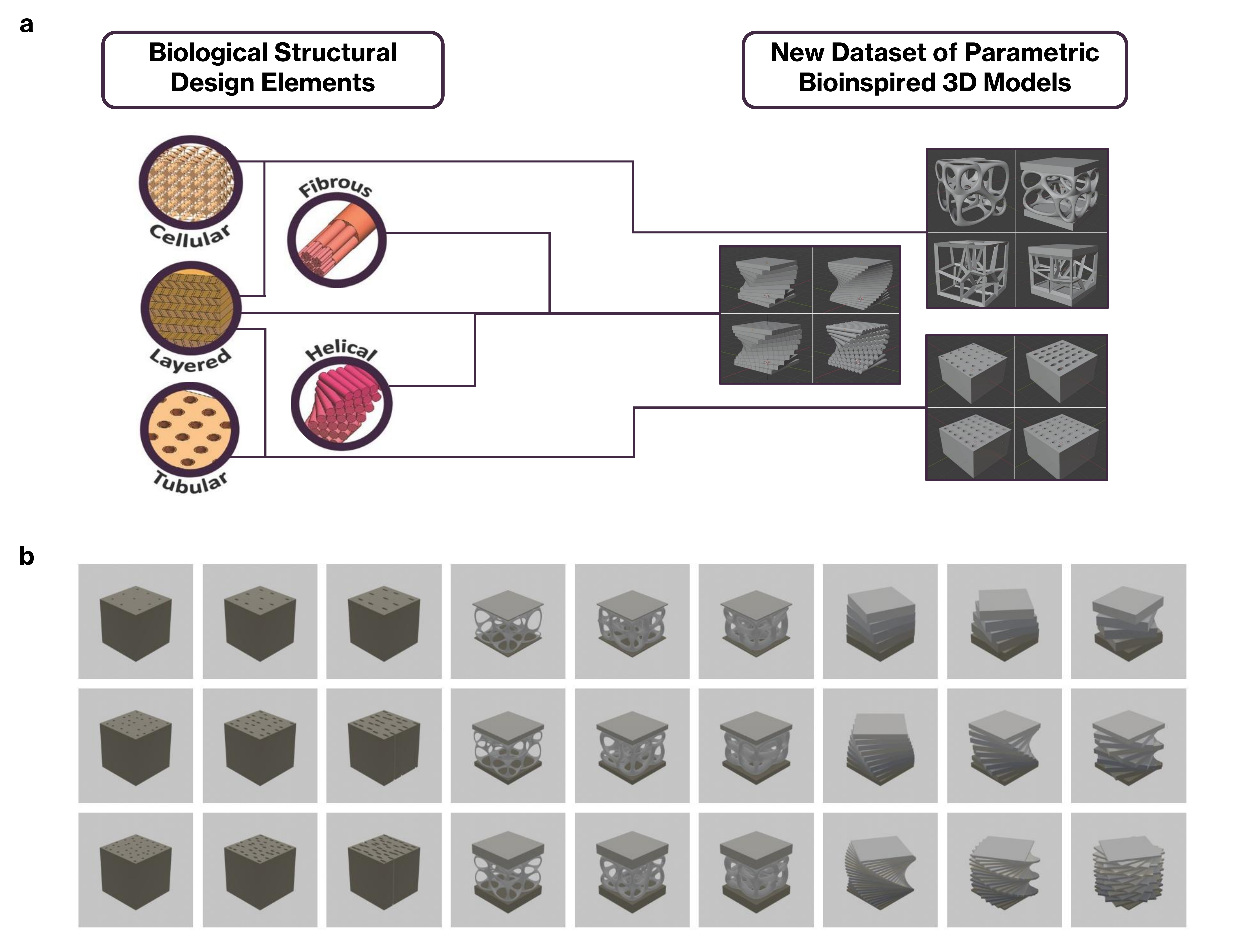}
    \caption{Bioinspired dataset overview. a) Common biological structural design elements, including cellular, fibrous, layered, helical, and tubular motifs, and their correspondence to the three parametric classes of bioinspired 3D structures introduced in this work. b) Representative examples of 3D structures illustrating variation across key geometric parameters for each class.}
    \label{fig:Fig_2}
\end{figure}

Figure \ref{fig:Fig_2}a showcases the biological design elements and their relevance to the curated dataset of 3D structures. Figure \ref{fig:Fig_2}b illustrates representative samples spanning the parameter space of each class, including tubular architectures varying in tubule count and ellipticity, cellular architectures varying in sandwich layer thickness and open cell dimensions, and helical architectures varying in ply count and rotation angle. While the dataset does not exhaustively cover all possible bioinspired geometries, it provides a structured foundation for capturing and exploring key classes of complex biological architectures.

\subsubsection{Cellular}

Cellular biological materials are widespread in nature and have been extensively studied in the context of cellular solids\cite{Gibson2005BiomechanicsSolids}, including bone\cite{Rowe2025IntegratingMechanics, Rho1998MechanicalBone, Olszta2007BonePerspective} and plant tissues\cite{Zhao2018OnPlants,Gibson2012TheMaterials,Speck2011PlantMechanics}, as well as in a wide range of biological systems such as bird beaks\cite{Seki2012StructureBeak,Lee2014HierarchicalBeak}, horseshoe crab shells \cite{Chen2012BiologicalDesigns}, and turtle shells \cite{Achrai2013Micro-structureShield}. These architectures are often accompanied by a sandwich layered configuration, in which a porous cellular core is enclosed by denser outer layers that contribute to stiffness, toughness, or impact resistance\cite{Haldar2014MechanicsCore}. To generate cellular structures in this dataset, controlled stochasticity is introduced to better capture biological variability. Cellular geometries are created using a fracture based approach, in which an initial solid volume is subdivided into a specified number of regions with randomized perturbations. This process is based on Voronoi partitioning \cite{Aurenhammer1991VoronoiStructure}, which divides space into regions defined by proximity to seed points and has been widely used to study cellular and foam like structures in biological materials\cite{Li2025Bio-inspiredTessellation,Lin2024Bio-inspiredBones,Tung2023OptimizationAlgorithm}. The number of fracture regions, degree of randomness, and resulting cell morphology are treated as tunable parameters. 

Another defining characteristic of biological cellular materials is the presence of smooth, curved interfaces rather than sharp, linear boundaries\cite{Naleway2015StructuralBioinspiration}. To introduce geometric smoothness, we apply Catmull Clark subdivision surfaces \cite{Qin1998DynamicSurfaces} to the generated meshes. This subdivision scheme recursively refines polygonal faces while adjusting vertex positions through local averaging, producing smoother surfaces at the cost of increased geometric resolution. The subdivision level is also treated as a controllable parameter.

\subsubsection{Helical}

Helical architectures are pervasive across biological length scales, ranging from nanoscale alpha helices to microscale and mesoscale structural arrangements. A prominent example is the Bouligand structure observed in the stomatopod dactyl club\cite{Weaver2012TheHammer}, where helicoidal stacking of fibrous layers contributes to exceptional impact resistance. More subtle helical motifs appear in sea sponges skeletal lattice \cite{Weaver2007HierarchicalAspergillum}, insect exoskeletons\cite{Lenau2008ColoursInterference}, and collagen organization in bone\cite{Miller1984Collagen:Bone, Buehler2006NatureFibrils}, where they contribute to in-plane isotropy and crack deflection\cite{Zimmermann2013MechanicalArmour, Lin2011MechanicalScales}. In this dataset, helical structures are parameterized by the number of layers, individual layer thickness, and the rotation angle between successive plies about a central axis. Fiber cross sectional geometry is also varied, allowing both cylindrical and rectangular arrangements. To better approximate biological irregularity, stochastic noise is introduced into the ply rotation, randomly perturbing the rotation angle between layers. This controlled randomness produces deviations from ideal helices while preserving the overall structural motif.

\subsubsection{Tubular}

The tubular class of bioinspired materials consists of elongated, aligned pores or tubules embedded within a surrounding bulk matrix. In natural systems, tubules vary in shape, size, density, orientation, and spatial arrangement, and are often accompanied by a denser cortical layer that surrounds the porous region and may possess distinct material properties. Tubular architectures are found in a range of biological materials, including the equine hoof wall \cite{Huang2019AWall,Lazarus2023EquineRevealed,Mahrous2023MultimoduleStructure}, bighorn ram horns\cite{Huang2017HierarchicalHorn,Zhang2018MicrostructureHorns}, crab exoskeletons\cite{Chen2008StructureExoskeletons}, and human tooth dentin\cite{Montoya2016ImportanceDentin,Kinney2003TheLiterature}. These structures are thought to contribute to energy absorption, crack deflection, and fracture control by guiding damage along preferred paths\cite{Lazarus2020AStructures,Feng2025EnergyStructures}. In the dataset, tubular structures are parameterized by tubule count, ellipticity, spacing, and the thickness of surrounding cortical layers, enabling systematic exploration of structure property relationships within this design class.

\subsection{Dataset Processing}
High level statistics of the final dataset are summarized in Table \ref{tab:bio3d-dataset}. The final dataset composition is detailed in Table \ref{tab:bio3d-instruction-types}, which outlines the different entry types and serves as a reference for the dataset components described in greater detail in the following sections.

\begin{table}[h]
\centering
\small
\begin{tabular}{l c}
\toprule
\textbf{Dataset Property} & \textbf{Value} \\
\midrule
Input modality & Natural language prompt \\
Output modality & Blender Python script (with or without embedded reasoning) \\
Samples & 4,558 \\
Average script length & $\sim$680 tokens \\
Total token count & $\sim$3.1M tokens \\
\bottomrule
\end{tabular}
\caption{High-level statistics for the Bioinspired3D fine-tuning dataset. Each sample consists of a natural-language prompt paired with a Blender Python script, accompanied by a rendered canonical view used for validation.}
\label{tab:bio3d-dataset}
\end{table}

\begin{table}[h]
\centering
\small
\begin{tabular}{l p{5.8cm} c c c c}
\toprule
\textbf{Type} & \textbf{Input Example} & \textbf{\# Entries} & \textbf{Pct.} & \textbf{Avg tokens} & \textbf{Range} \\
\midrule
General Blender Manipulations 
& ``Write Blender Python script that constructs a 3x3 grid of cubes each rotated slightly in Z'' 
& 727 & $\sim16\%$ & 129.2 & 48--435 \\

Bioinspired 
& ``generate a Blender script that builds a helical twisted ply structure'' 
& 827 & $\sim18\%$ & 445.9 & 177--1016 \\

Bioinspired + Reasoning 
& ``Write Blender script that makes a tubular porous material. Think step by step'' 
& 933 & $\sim20\%$ & 1253.4 & 691--2252 \\

BlendNet 
& ``blender script to create a model of a dining table'' 
& 1029 & $\sim23\%$ & 402.9 & 154--1298 \\

BlendNet + Reasoning 
& ``with python blender script make a sofa. think step by step'' 
& 1041 & $\sim23\%$ & 1007.0 & 437--2843 \\
\midrule
\textbf{Total} &  & \textbf{4558} & \textbf{100\%} &  &  \\
\bottomrule
\end{tabular}
\caption{Composition and token statistics for the Bioinspired3D training corpus. The dataset spans five instruction types, from simple geometric manipulations to complex bioinspired reasoning tasks.}
\label{tab:bio3d-instruction-types}
\end{table}

\subsubsection{Bioinspired Entries}

To prepare the bioinspired 3D dataset for supervised finetuning, we developed a dataset processing pipeline to expand an initial set of base scripts into a sufficiently large and diverse training corpus. The original dataset consisted of 12 manually designed base scripts, with four scripts per bioinspired class. To scale the dataset while limiting overfitting, we intentionally diversify how scripts are written for supervised fine-tuning. Because scripting tasks rarely have a single correct solution, we include multiple structurally equivalent but syntactically distinct representations.

\begin{figure}[h]
    \centering
    \includegraphics[width=1\linewidth]{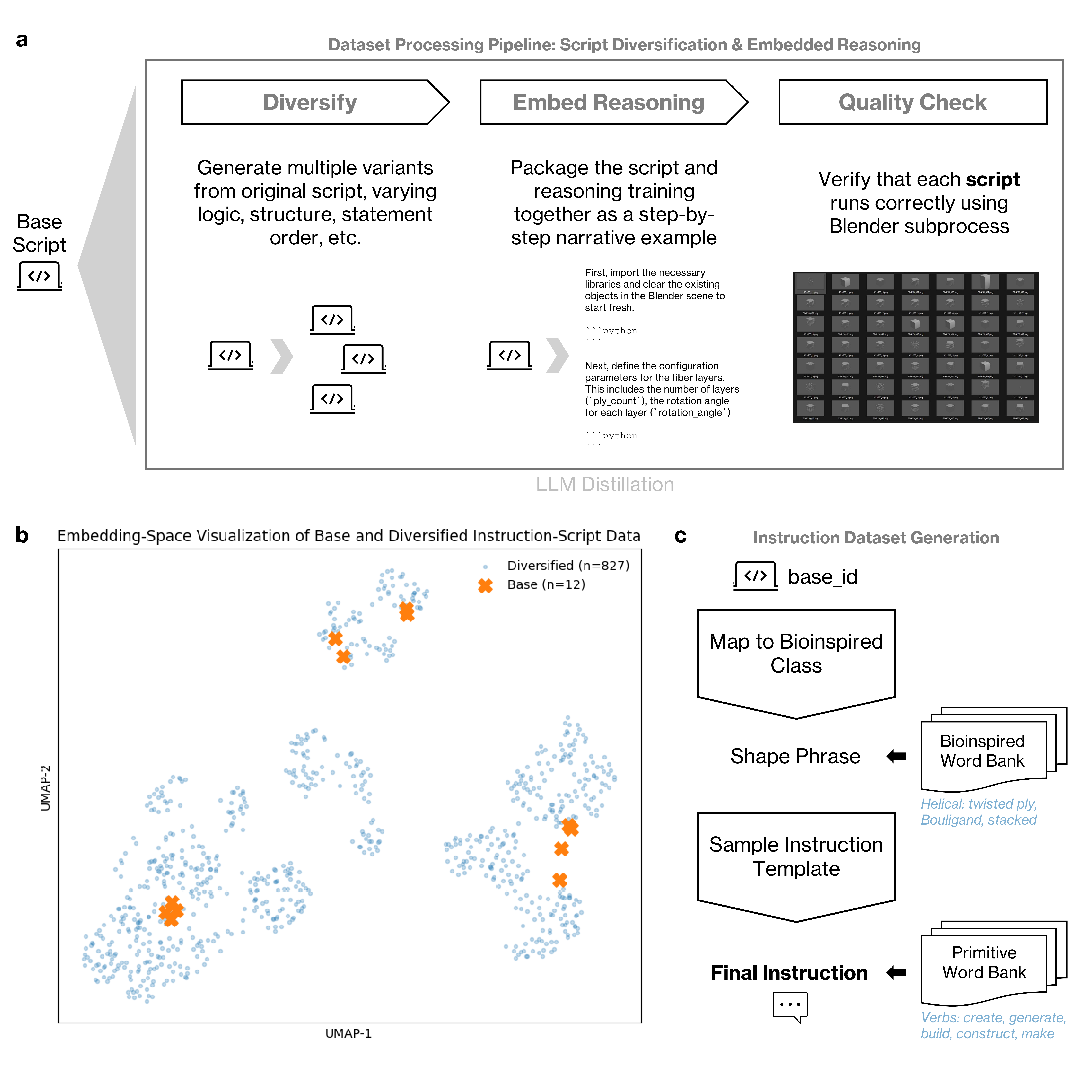}
    \caption{a) Data processing pipeline powered by LLM distillation, consisting of three phases: diversification of base scripts into multiple coherent variants, embedded reasoning in which each script is packaged as a narrative step by step example, and validation via headless Blender subprocess execution to ensure successful runtime and correct geometry based on visual inspection of rendered outputs. b) Embedding space visualization of base and diversified instruction script pairs generated through the dataset pipeline. c) Instruction dataset generation process. For each script, a base identifier is retrieved to determine the corresponding bioinspired material class (helical, tubular, or cellular). A shape phrase is constructed by sampling from a class specific Bioinspired Word Bank, which is then combined with a randomly sampled instruction template and a primitive word bank containing varied verbs and grammatical forms to generate natural language instructions.}
    \label{fig:Fig_3}
\end{figure}

The dataset processing pipeline is illustrated in Figure \ref{fig:Fig_3}a. Inspired by prior work on LLM distillation\cite{Luu2023BioinspiredLLM:Materials, Lu2024Fine-tuningCapabilities}, we use GPT-4o-mini\cite{achiam2023gpt} as a teacher model to generate new script variants and embed structured reasoning. The pipeline consists of two main stages: diversification and embedded reasoning.

In the first stage, diversification, the model is prompted with one of the original base Blender Python scripts and instructed to generate multiple functionally equivalent variants by altering parameters, code structure, and implementation style while preserving the overall structural and geometric outcome. For each base script, the model is sampled multiple times, producing five variants per query to expand coverage of the design space.

In the second stage, embedded reasoning, each diversified script is independently passed back to the model and reformatted into a narrative, step by step explanation of the generation process. The model is provided with an example narrative structure in which prose explanations are interleaved with Python script blocks, and the complete executable script is returned at the end. This step embeds explicit reasoning into the dataset while preserving a clean separation between natural language descriptions and executable script.

All generated scripts, including both diversified and reasoning-enriched variants, are automatically extracted and passed to a headless Blender subprocess for validation. Each script is executed to generate geometry, after which the scene is automatically configured with camera placement and lighting and then rendered. The rendered outputs are visually inspected for quality control to ensure that, while parameter values may vary, the resulting geometry still reflects the intended bioinspired design motif.

To assess the impact of the diversification step, both the original base scripts and the diversified variants are embedded into a shared representation space and visualized using UMAP, as shown in Figure \ref{fig:Fig_3}b. The axes of the UMAP projection are arbitrary and distances should be interpreted qualitatively. Nevertheless, the visualization reveals clear trends, with diversified script variants spreading across the embedding space and diverging from the original 12 base script clusters. This indicates that the diversification process effectively expands the representation space beyond a small set of isolated points, increasing coverage and variability in the training data. Although diversification preserves the same geometric outcome, showing the model multiple ways to implement a design helps it learn general construction patterns rather than memorize fixed templates, enabling better generalization to new designs at inference time.

For input prompt generation, we employ a template based algorithm designed to emulate the variability of natural language instructions. A set of instruction templates with diverse grammatical structures is designed to capture differences in phrasing, capitalization, and sentence flow. Two complementary word banks are used in this process. The first is a bioinspired motif word bank, which contains multiple semantically equivalent terms describing each bioinspired concept. The second is a primitive word bank, consisting of verbs, mediums, and action words across grammatical tenses. As illustrated in Figure \ref{fig:Fig_3}c, each script is first mapped to its corresponding bioinspired material class. Terms are then randomly sampled from the bioinspired motif word bank and inserted into a selected instruction template. Primitive words are sampled and slotted into the remaining template fields to construct a final natural language instruction. This procedure produces diverse natural prompts while preserving a consistent mapping between instruction semantics and the underlying 3D geometry.

\subsubsection{General Blender Entries}
To imbue the final model with general knowledge of Blender operations while further increasing dataset diversity, we generate additional data focused on a set of generic 3D design tasks. These tasks span three levels of complexity. The first involves primitive shape generation with non-default parameters. The second includes geometric transformations such as rotation about an axis, orientation, translation in the x, y, and z directions, grid and radial layouts, and axis-specific scaling. The third covers more advanced operations, including Boolean difference operations, relative object placement, and loop-based generation of multiple objects.

The same diversification pipeline used for the bioinspired entries is applied here for LLM distillation. For each task, the model is prompted to solve the task in multiple distinct ways, with unique solutions generated per query and sampled repeatedly to produce additional variants. All successfully generated scripts are executed and subsequently quality checked through visual inspection of the rendered outputs to ensure correctness.

\subsubsection{BlendNet Entries}
Finally, to further expand the model’s general Blender knowledge beyond bioinspired geometries, we incorporate a curated subset of the BlendNet synthetic dataset introduced by BlenderLLM\cite{Du2024BlenderLLM:Self-improvement}. We first visually inspect the full dataset and select 380 high quality examples of well constructed Blender objects, excluding samples with floating components, self intersections, or inconsistent materials. These selected samples are then passed through the same diversification and embedded reasoning pipeline used for the bioinspired entries. All resulting variants are executed and quality checked to ensure geometric validity and consistency.

\subsection{Bioinspired3D - Finetuned Model}
With the completed dataset, we finetune the base model Llama-3.2-3B-Instruct using a low-rank adaptation (LoRA) strategy \cite{Hu2021LoRA:Models}. This approach freezes the base model weights while introducing a small number of trainable adapter layers. Full training hyperparameters are reported in Table \ref{tab:bio3d-hparams}. Detailed training dynamics are included in the Supplementary Information.

\begin{table}[h]
\centering
\small
\begin{tabular}{l c}
\toprule
\textbf{Hyperparameter} & \textbf{Value} \\
\midrule
Base model & Llama-3.2-3B-Instruct \\
LoRA rank ($r$) & 64 \\
LoRA $\alpha$ & 64 \\
LoRA dropout & 0.10 \\
Target modules & q, k, v, o, up, down, gate \\
Precision & FP16 \\
\midrule
Per-device train batch size & 1 \\
Gradient accumulation steps & 8 \\
Effective batch size & 8 \\
Learning rate & $1\times10^{-4}$ \\
Warmup steps & 50 \\
LR scheduler & Linear decay \\
Optimizer & AdamW \\
Adam betas & (0.9, 0.999) \\
Adam epsilon & $1\times10^{-8}$ \\
Weight decay & 0.0 \\
Max gradient norm & 1.0 \\
Seed & 42 \\
\bottomrule
\end{tabular}
\caption{Fine-tuning hyperparameters for the Bioinspired3D model, finetuned Llama-3.2-3B-Instruct.}
\label{tab:bio3d-hparams}
\end{table}

Model performance is evaluated using a custom benchmark composed of novel natural language prompts designed to test bioinspired 3D generation. Evaluation proceeds in two stages. First, we check whether the generated script runs successfully in Blender to assess syntactic and functional validity. This check is performed automatically using a headless Blender subprocess.

Secondly, model outputs are evaluated based on rendered geometry. Generated scripts are extracted, executed, and rendered, and the resulting images are exported for visual inspection. Because the benchmark emphasizes difficult prompts and previously unseen structures, grading is performed comparatively rather than against a fixed ground truth. Outputs are scored on a continuous scale from 0 to 1, with partial credit awarded for plausible structural interpretations even when generation is imperfect.

The final finetuned model checkpoint is then equipped with retrieval augmented generation (RAG)\cite{Lewis2020Retrieval-AugmentedTasks}. In this setting, the retrieval database consists of a JSON file containing concise linguistic descriptions of the 12 original bioinspired structures along with their corresponding base scripts. 

The final finetuned model configuration was selected based on a unified ablation study examining retrieval depth, dataset composition, and sampling temperature as shown in Table \ref{tab:bio3d-ablation}.  Across retrieval settings, moderate retrieval depth provided the strongest performance, with retrieval augmented generation using $k=2$ where $k$ denotes the number of retrieved context entries included in the prompt. This choice provides sufficient structural guidance from prior examples while avoiding prompt overcrowding that can disrupt coherent script generation. Increasing retrieval beyond this point led to diminished returns, suggesting that excessive contextual information introduces noise rather than additional useful signal for 3D script generation. Qualitative examples exploring the effects of retrieval are documented in Supplementary Information. 

Dataset composition played a significant role in overall performance. Incorporating retrieval augmented generation consistently improved results compared to finetuning alone, while removing either embedded reasoning variants or the BlendNet subset resulted in noticeable performance degradation. This indicates that both narrative reasoning and exposure to general Blender object structures contribute meaningfully to the model’s ability to generate valid and interpretable 3D scripts. An example of outputs with and without reasoning are provided and analyzed in the Supplementary Information. 

Finally, sampling temperature influenced generation quality, with low temperature yielding the most stable and accurate outputs, particularly when combined with retrieval augmentation. This trend is expected, as the model was finetuned using low temperature generations, biasing it toward deterministic decoding during inference. Nevertheless, performance degrades noticeably at higher temperatures, especially in the absence of retrieval. Based on these findings, the final Bioinspired3D configuration uses RAG with $k=2$, incorporates both reasoning enriched and BlendNet data, and operates at a low sampling temperature.

\begin{table}[h]
\centering
\small
\begin{tabular}{l c}
\toprule
\textbf{Ablation Study} & \textbf{Score} \\
\midrule
\multicolumn{2}{l}{\textbf{RAG (varying k)}} \\
\midrule
k = 1 & 0.566 \\
k = 2 & \textbf{\underline{0.600}} \\
k = 3 & 0.547 \\
\midrule
\multicolumn{2}{l}{\textbf{Dataset Composition}} \\
\midrule
Bio3D (no RAG)              & 0.542 \\
Bio3D (with RAG)            & \textbf{\underline{0.600}} \\
No Reasoning (with RAG)     & 0.528 \\
No BlendNet (with RAG)      & 0.520 \\
\midrule
\multicolumn{2}{l}{\textbf{Temperature}} \\
\midrule
t = 0.1 (with RAG) & \textbf{\underline{0.600}} \\
t = 0.1 (no RAG)          & 0.542 \\
t = 0.5 (with RAG)          & 0.583 \\
t = 0.5 (no RAG)            & 0.505 \\
\bottomrule
\end{tabular}
\caption{Unified ablation study covering retrieval depth (RAG-$k$), dataset composition, and sampling temperature. RAG at $k{=}2$, inclusion of both reasoning and BlendNet data, and a low sampling temperature ($t{=}0.1$) yield the strongest performance.}
\label{tab:bio3d-ablation}
\end{table}

More important than Blender validation success is performance on the benchmark itself, namely whether the model can design novel 3D structures when prompted. Benchmark correctness results for both the base and finetuned models are shown in Figure \ref{fig:Fig_5}a. The base model alone achieves a score of 0, indicating an inability to generate meaningful 3D bioinspired designs. When equipped with retrieval augmented generation, its performance improves modestly to 18\%. In contrast, the finetuned model achieves a score of 54\%, which further increases to 60\% when combined with retrieval. This represents more than a three fold improvement over the base model with retrieval, demonstrating that targeted finetuning can effectively teach foundational models new, structured generation capabilities.

Figure \ref{fig:Fig_5}b breaks down performance by question difficulty. The finetuned models perform consistently well on easy and medium difficulty prompts, while performance decreases on hard prompts, which constitute the majority of the benchmark. This trend is expected, as hard prompts often require the combination of multiple bioinspired motifs or large extrapolation beyond patterns explicitly observed during training. In contrast, the base model performs poorly across all difficulty levels, indicating that retrieval alone is insufficient to enable meaningful 3D design behavior.

Although the easy and medium prompts remain distinct from the training data, they are more closely aligned with learned structures, differing primarily in phrasing or the inclusion of novel constraints such as explicit numerical parameter values. Figure \ref{fig:Fig_5}c highlights representative examples of successful hard prompt generations, including designs that combine motifs across different bioinspired classes. These examples demonstrate the model’s ability to synthesize and manipulate new structural concepts that were not explicitly present in the training set.

\subsection{Bioinspired123D - Agentic System}

To further improve the performance of Bioinspired3D, we analyzed benchmark failures and identified two dominant failure modes. The first consists of minor script errors, which occur infrequently. The second, and more significant, failure mode arises from incorrect or incomplete model generation, particularly for hard level prompts that require multi motif reasoning or extrapolation beyond the training distribution. To address these limitations, we extend Bioinspired3D into an agentic framework in which the model is no longer a standalone prompt to script generator, but instead operates as part of a coordinated system designed to support iterative reasoning and correction.

The resulting graph based agentic system embeds Bioinspired3D within a set of lightweight, task specific agents. In this framework, agents are represented as nodes, and edges define transitions between agents based on the current design state. All agents operate on shared state variables, including generated script, rendered geometry, and intermediate evaluation results. Notably, all agents are off the shelf and lightweight, requiring no additional finetuning and minimal computational overhead.

\begin{figure}[h]
    \centering
    \includegraphics[width=1\linewidth]{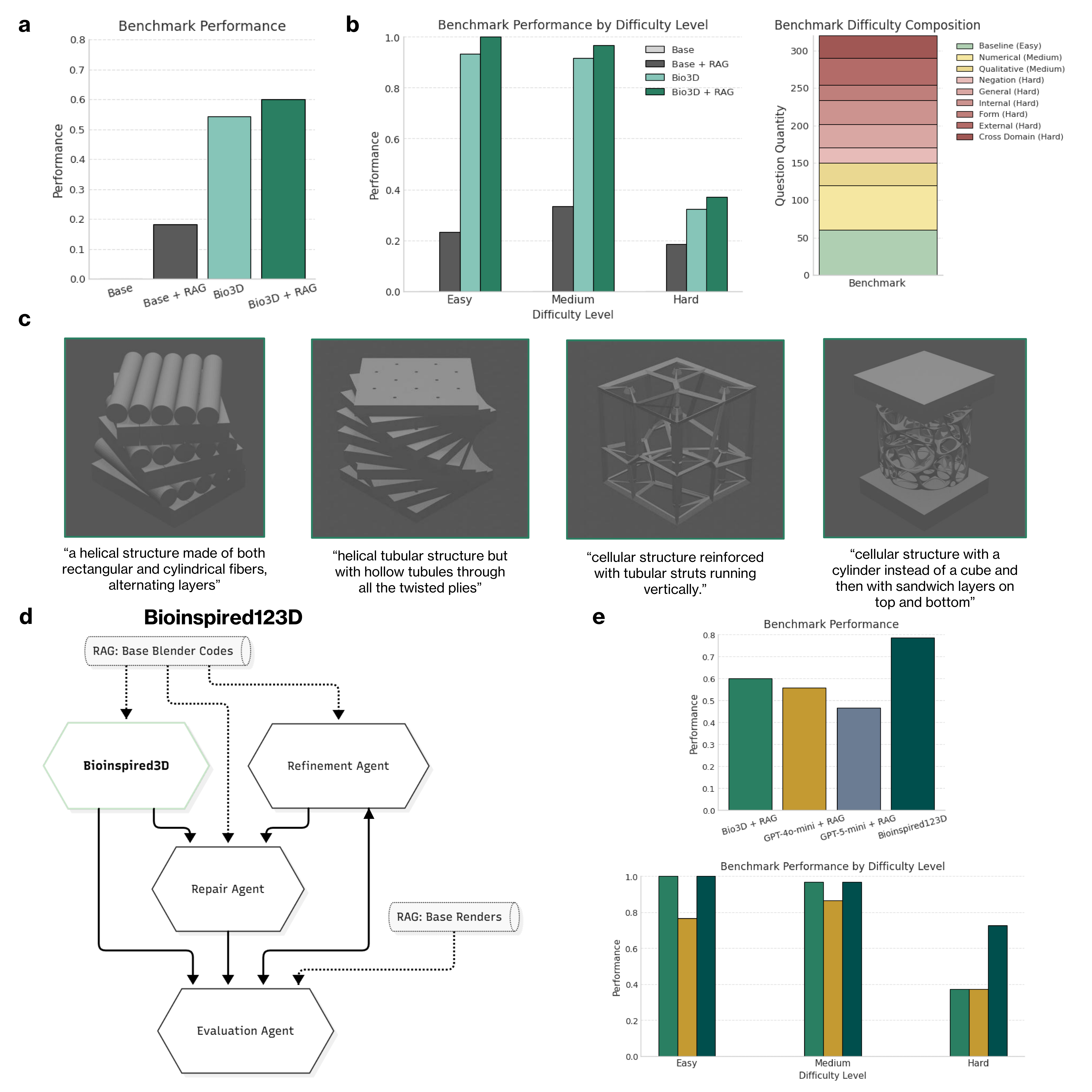}
    \caption{a) Benchmark performance across models, including the base model, the final finetuned checkpoint Bioinspired3D, with and without RAG. b) Benchmark performance across models, separated by benchmark question difficulty level, along with a stacked bar plot displaying the benchmark composition showing strong concentration of "hard" leveled questions. c) Examples of successful generations for "hard" level questions in which the prompts utilize both new phrasing and request creative extrapolated ideas that go beyond the dataset. d) Depiction of the Bioinspired123D graph-based agentic system. e) Benchmark performance across models, now including Bioinspired123D and state of the art models, GPT-4o-mini and GPT-5-mini, with breakdown by difficulty level.}
    \label{fig:Fig_5}
\end{figure}

Two auxiliary language model agents are introduced: a Repair Agent, which focuses on correcting syntactic or execution level errors, and a Refinement Agent, which improves geometric fidelity and adherence to the prompt intent. In addition, a small vision language model (VLM) is incorporated as an Evaluation Agent, responsible for assessing rendered outputs. All language model agents, including Bioinspired3D, share access to the same retrieval augmented database containing bioinspired base scripts. The Evaluation Agent is provided with a separate retrieval database consisting of image paths to rendered examples of base bioinspired structures. The complete system architecture and interaction flow are shown in Figure \ref{fig:Fig_5}d, where agent transitions are governed by predefined design state conditions.

Figure \ref{fig:Fig_5}e compares the performance of the full agentic system, referred to as Bioinspired123D, against Bioinspired3D alone as well as other state of the art foundation models equipped with retrieval augmentation, including GPT-4o-mini and GPT-5-mini. While the exact parameter counts of these proprietary models are not disclosed, they are assumed to be substantially larger than the 3B parameter base model used in this work. Despite this, Bioinspired3D alone outperforms the mini GPT models across easy and medium prompts and achieves comparable performance on hard prompts.

When embedded within the agentic system, Bioinspired123D demonstrates further gains, particularly on hard level questions. Notably, performance improvements on hard prompts exceed what would be expected from simple averaging between Bioinspired3D and GPT-4o-mini. Instead, results exhibit a performance spike, suggesting the emergence of complementary behavior arising from structured agent interactions.

The improved performance of the integrated system is largely driven by the interaction between the Evaluation and Refinement agents. The vision language model serves as an Evaluation Agent by analyzing rendered outputs and comparing them against the original design prompt. Although the VLM is not explicitly trained on bioinspired structures, it is provided with reference renders of base bioinspired structures through retrieval augmentation, enabling relative comparison rather than absolute classification.

The VLM performs two primary checks. First, it identifies physical inconsistencies such as overlapping geometry, floating components, or artifacts. Second, it evaluates semantic correctness by assessing whether the rendered structure aligns with the design intent expressed in the prompt. Based on this assessment, the VLM issues feedback that is passed to the Refinement Agent, which modifies the underlying Blender script accordingly. This process is repeated for a limited number of iterations, after which the VLM performs a final acceptance decision.

\begin{figure}[h]
    \centering
    \includegraphics[width=1\linewidth]{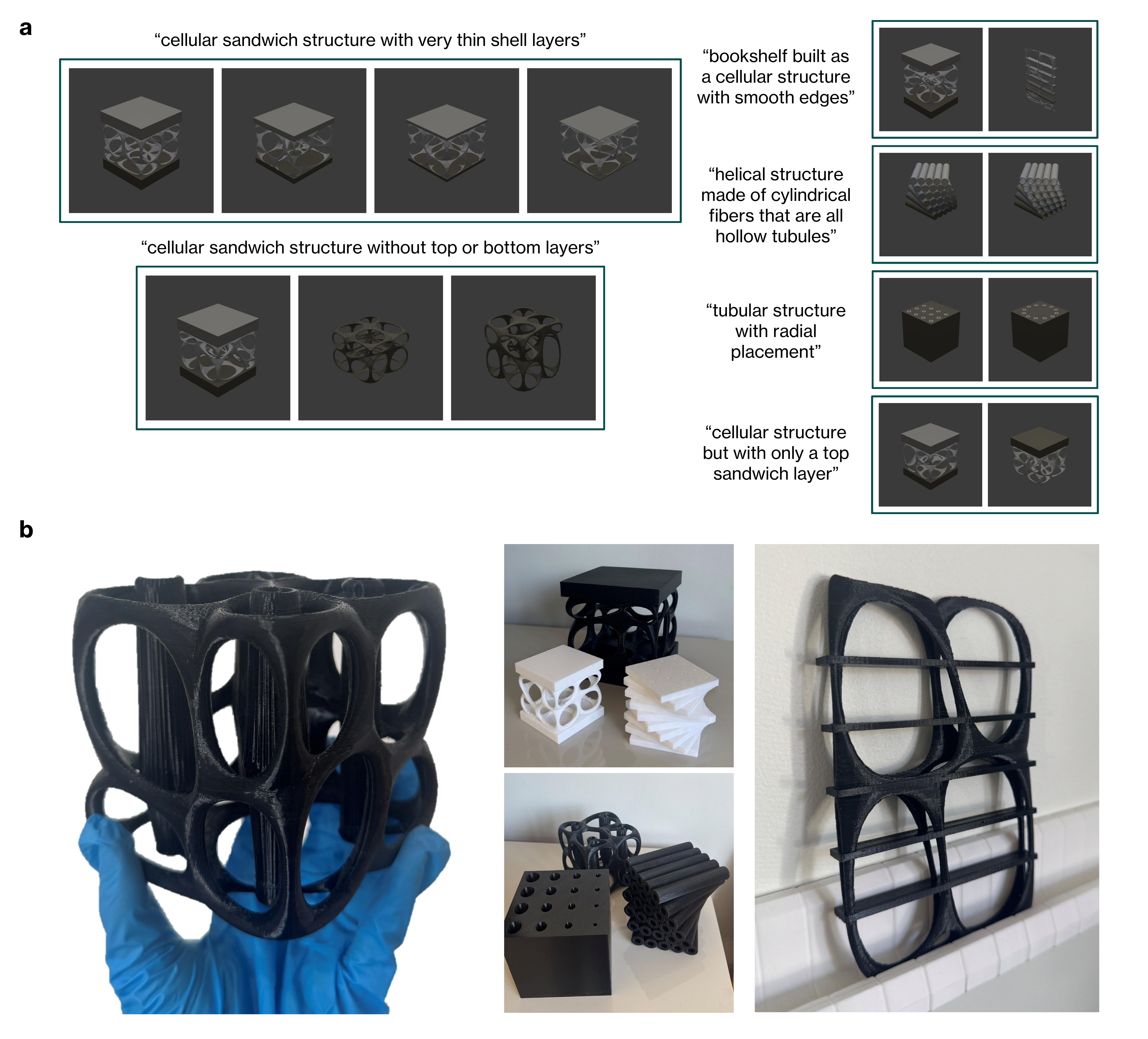}
    \caption{a) Examples of evaluation prompts passed to Bioinspired123D with progressive renders from the agent system, refining the structures chronologically from left to right. b) Photographs of 3D printed samples of the novel AI generated 3D structures. }
    \label{fig:Fig_6}
\end{figure}

Figure \ref{fig:Fig_6}a illustrates this iterative refinement process for representative hard level prompts that previously would have terminated after a single generation attempt. In one example involving a cellular sandwich structure with thin shell layers, successive iterations progressively reduce shell thickness, culminating in a final design after four refinement steps, which corresponds to the maximum number of allowed attempts. In other cases, the VLM identifies overly generic initial generations, and its feedback enables the Refinement Agent to steer the design toward a more faithful interpretation of the prompt. As illustrated in Figure \ref{fig:Fig_6}a, a prompt requesting a bookshelf with a smoothed cellular motif fails to capture the shelf structure, instead producing a generic cellular geometry due to semantic drift toward an in-distribution design. Through iterative VLM feedback and refinement, the system course-corrects toward a stretched cellular structure augmented with flat slab-like elements that serve as shelves. This example highlights both the flexibility and the limits of out-of-distribution generalization. While the system can reuse learned geometric primitives to approximate familiar concepts such as a bookshelf, the extent of this extrapolation remains bounded by the knowledge encoded in the agents.

While the VLM is not perfect and can be conservative in certain cases, these results demonstrate that iterative visual feedback plays a critical role in correcting complex failure modes that are difficult to address through single pass generation alone.

To select an appropriate vision–language model for the Evaluation Agent within Bioinspired123D, we evaluate off-the-shelf VLMs under different retrieval and scoring settings, including GPT-4o-mini and Qwen-3-VL-2B-Instruct \cite{Bai2025Qwen3-VLReport}. The Evaluation Agent plays a critical role in the agentic pipeline by assessing rendered outputs for both physical validity and semantic alignment with the design prompt. As such, its performance directly affects downstream refinement and overall system behavior.

We consider two evaluation modes to probe the behavior of the Evaluation Agent. In "any step" evaluation, the highest scoring render across the full generation sequence is selected. This mode captures cases in which the model produces a correct or high quality design at an intermediate iteration, even if subsequent refinement attempts is incorrect. In contrast, the final render evaluation considers only the last generated output, reflecting scenarios in which the Evaluation Agent enforces continued refinement until a terminal step is reached.

Comparing these two modes highlights a known failure mode of vision language model based evaluation, in which an overly conservative evaluator may reject an already valid design and trigger further refinement that ultimately worsens the output. In practical usage, Bioinspired123D provides all intermediate scripts and renders to the user, allowing manual selection of the most suitable result. As such, the any step evaluation more closely reflects realistic user workflows, while the final render evaluation provides insight into the strictness and stability of the automated evaluation loop.

The results are presented in Table \ref{tab:model-rag-final-ablation}. Across both evaluation modes, the GPT based VLM consistently outperform the Qwen based VLM, indicating stronger multimodal reasoning and visual grounding. For GPT, retrieval augmentation improves performance in both settings, with the strongest results observed under any step evaluation. This suggests that retrieval provides useful contextual priors that help the evaluator correctly identify high quality intermediate designs, even in cases where subsequent refinement steps lead to degradation. 

\begin{table}[h]
\centering
\small
\begin{tabular}{l c c c c}
\toprule
\textbf{Evaluation Mode} 
& \textbf{GPT (no RAG)} 
& \textbf{GPT (RAG)} 
& \textbf{Qwen (no RAG)} 
& \textbf{Qwen (RAG)} \\
\midrule
Any-step (best of sequence) 
& 0.7578 & \textbf{\underline{0.7688}} & 0.7109 & 0.6859 \\
Final render only 
& 0.6234 & 0.6859 & 0.5172 & 0.6516 \\
\bottomrule
\end{tabular}
\caption{VLM Comparison, comparing model family, RAG usage, and evaluation scope. Any-step scoring takes the best render from the full generation sequence. Final render scoring evaluates only the last frame.}
\label{tab:model-rag-final-ablation}
\end{table}

Given the reliance on VLMs, particularly proprietary systems such as GPT based models, we examine the system level tradeoffs associated with their use as shown in Table \ref{tab:vlm-token-and-system-comparison}. Retrieval augmented evaluation more than doubles total token consumption, primarily due to additional image inputs. While this overhead yields clear performance gains for stronger VLMs, it introduces nontrivial cost and latency considerations that must be accounted for in large scale or fully automated deployments.

We further compare local and API based deployment options for the Evaluation Agent. Local models such as Qwen-VL offer zero marginal inference cost and full control over hardware and data, but underperform on complex multimodal judgments. In contrast, API based models such as GPT-VLM provide stronger evaluation accuracy and require minimal local setup, at the expense of usage based pricing and external dependencies. These results suggest that hybrid evaluation strategies, in which local VLMs handle routine evaluations and API based models are reserved for challenging cases, can offer a practical balance between performance, cost, and scalability.

Together, these findings inform the design of the Evaluation Agent in Bioinspired123D and underscore the importance of aligning VLM choice, retrieval strategy, and evaluation protocol with both system constraints and realistic user workflows.

\begin{table}[h]
\centering
\small
\begin{tabular}{lccc}
\toprule
\textbf{Scenario} & \textbf{Text Tokens} & \textbf{Images} & \textbf{Total Tokens} \\
\midrule
No RAG 
& $\sim$300--350 
& 1 image ($\sim$600 eq.) 
& $\sim$900--1{,}000 \\
RAG (k=2) 
& $\sim$360--400 
& 3 images ($\sim$1{,}800 eq.) 
& $\sim$2{,}100--2{,}300 \\
\bottomrule
\end{tabular}

\vspace{8pt}
\begin{tabular}{l p{3.8cm} p{3.8cm}}
\toprule
\textbf{Category} & \textbf{Qwen-VL (Local)} & \textbf{GPT-VLM (API)} \\
\midrule
Cost 
& Free (open-source, local inference) 
& Usage-based pricing (e.g., \$0.15 / 1M input tokens) \\
\midrule
Inference setup 
& Fully local inference; model loaded in FP16 
& Hosted inference via OpenAI API; no local weights \\
\midrule
Hardware requirements 
& Requires GPU VRAM (2--8 GB) 
& None (compute offloaded to API servers) \\
\bottomrule
\end{tabular}

\caption{
\textbf{Top:} Estimated input-token costs for the automated rendering-evaluation pipeline, with and without RAG retrieval. 
GPT-4o-mini does not use separate image-token pricing; multimodal inputs are billed as standard input tokens.
\textbf{Bottom:} Comparison between Qwen-VL (local inference) and GPT-based vision-language models (API). 
}
\label{tab:vlm-token-and-system-comparison}
\end{table}

To further demonstrate the versatility of using Blender as the underlying framework, selected designs generated by Bioinspired3D and Bioinspired123D were exported from Blender as STL files and fabricated via 3D printing. Representative printed samples are shown in Figure \ref{fig:Fig_6}b, illustrating that the generated geometries are directly compatible with downstream fabrication workflows. While these examples demonstrate fabricability, printer-specific constraints such as minimum feature size, overhang limits, and build-volume restrictions are not explicitly enforced during generation. However, the agent framework could readily support the inclusion of automated manufacturability checks within the loop. In practice, mesh-level checks performed after each Blender execution could identify violations of fabrication limits, which would be used by the Refinement Agent to adjust geometric parameters in subsequent iterations.

\begin{figure}[h]
    \centering
    \includegraphics[width=1\linewidth]{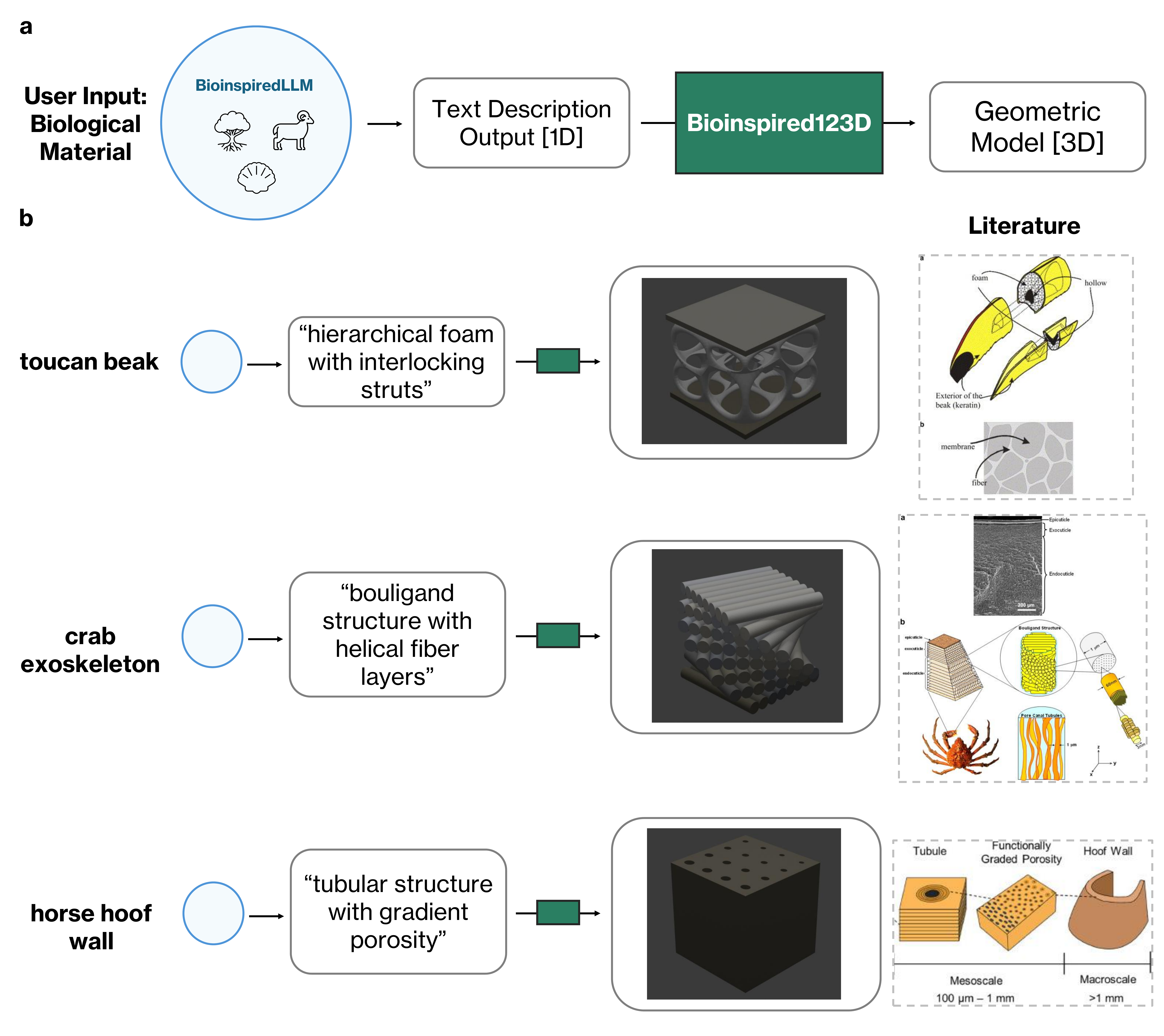}
    \caption{a) Translating from text description output 1D to geometric structures 3D by coupling BioinspiredLLM with the Bioinspired123D system, in which the initial user input is the name of a biological material. b) Example outputs from the full pipeline, including user inputs for "toucan beak", "crab exoskeleton", and "horse hoof wall" along with their BioinspiredLLM 1D generated prompts, the corresponding Bioinspired123D 3D structure and compared to schematic figures found in the literature. Literature images sourced from: toucan beak \cite{Seki2005StructureBeak} (Reproduced with permission from Elsevier © 2005), crab exoskeleton\cite{Chen2008StructureExoskeletons} (Reproduced with permission from Elsevier © 2008), horse hoof wall (Reproduced from Ref \cite{Bonney2024ViscoelasticWall}. Licensed under CC-BY.)}
    \label{fig:Fig_7}
\end{figure}

Finally, we demonstrate full integration of the system with its original design goal of coupling biological knowledge to mechanistically informed physical design. As shown in Figure \ref{fig:Fig_7}a, BioinspiredLLM is interfaced with Bioinspired123D in an end to end workflow that begins with a user provided biological material name. BioinspiredLLM reflects on the material’s known microstructural features and generates a concise textual design prompt. This prompt is then passed to Bioinspired123D, which produces a corresponding 3D geometric structure.

Figure \ref{fig:Fig_7}b presents representative examples spanning biological materials such as toucan beak, crab exoskeleton, and horse hoof wall. In each case, BioinspiredLLM generates a distinct design prompt describing structural features that Bioinspired123D was not explicitly trained on. Nevertheless, Bioinspired123D successfully interprets these linguistic cues and generates a 3D structure that qualitatively aligns with known structural schematics reported in prior literature. Notably, in the horse hoof wall example, the generated prompt describes a "tubular structure with gradient porosity", which Bioinspired123D interprets as a spatial variation in tubule size along a principal material axis. This behavior emerges despite the absence of explicit supervision on gradient-based architectures, one of the eight common design elements, in the training dataset, indicating that the system can infer and operationalize simple graded design concepts directly from natural language. These results demonstrate that the combined system can translate high level biological knowledge into plausible, mechanistically informed 3D designs without explicit supervision on specific biological instances.

\section{Conclusions}
In this work, we demonstrate that Bioinspired123D translates natural language design intent into physical 3D structures through a text to script pipeline that produces executable geometry. In contrast to text-to-mesh or diffusion-based 3D generation methods, Bioinspired123D represents geometry as compact, interpretable programs executed directly in Blender. This code-as-geometry approach enables precise structural control, low compute requirements, and direct compatibility with downstream simulation and fabrication workflows, without reliance on large-scale 3D foundation models.

A second contribution is an execution- and render-grounded refinement loop: generated scripts are executed in Blender, rendered, and then critiqued by a vision–language model whose feedback guides iterative script repair and refinement. This closes the loop between natural language intent, executable code, and observed geometry, improving prompt fidelity and robustness in cases where single-pass generation fails.

A key insight from this work is that much of the capability required for structured 3D design already exists within open-source, code-capable language models, in the form of embedded programming knowledge and natural language fluency. We find that models large enough to reliably generate and modify short procedural programs, but far smaller than contemporary multimodal or 3D foundation models, are sufficient when generation is grounded in executable code. By leveraging Blender’s Python interface, these capabilities can be accessed through a compact, domain-specific model without requiring large-scale geometric pretraining. The integration of agentic refinement further shows that contextual feedback over rendered outputs can correct minor errors, resolve ambiguities, and improve geometric fidelity without increasing model size or additional finetuning.

This work also highlights several open challenges. Unlike many generative tasks, bioinspired 3D design does not admit a single correct answer. Generated structures are often novel or creative, which complicates evaluation and benchmarking. While we introduce a custom benchmark and relative grading scheme, defining objective, scalable metrics for assessing correctness and quality remains an open problem.

Dataset quality presents a related challenge. While automated evaluation could accelerate dataset expansion, we intentionally prioritize execution-based validation and careful curation to ensure the long-term utility of the released dataset. Future work may explore tighter integration of vision–language models to partially automate filtering and grading, enabling larger datasets while maintaining quality. Similarly, using Bioinspired123D itself as a tool for controlled 3D data generation is a promising direction.

Looking forward, the modular design of Bioinspired123D enables straightforward extension. Future work could expand the dataset of bioinspired structural motifs beyond helical, cellular, and tubular families emphasized here to include multiscale and hierarchical motifs. These broader classes can be expressed in Blender scripts by composing motifs at multiple length scales (for example, instancing a smaller cellular unit within a larger scaffold) and by allowing key parameters to vary across space (for example, thickness or density as a function of position) to explicitly capture graded or anisotropic behavior.

The same agentic framework also naturally supports the integration of additional domain-specific critics. As discussed earlier, automated manufacturing checks for 3D printing could be incorporated into the loop. In a similar manner, a lightweight finite element analysis agent could provide coarse mechanical response estimates (for example, stiffness, compliance, or strain localization) that inform subsequent geometric refinement. Together, these extensions illustrate how domain knowledge can be incrementally layered onto a compact generative core. 

More broadly, this work demonstrates that scientific 3D design tools can be built using compact, modular systems that run on modest computational resources, helping broaden access to generative AI methods for materials and structural design.

\section{Materials and methods}
\label{sec:Materials_Methods}
 
We provide details on the materials and methods used to conduct this study. 

\subsection{Blender}
Blender \url{(https://www.blender.org/)} (Vers. 4.2 LTS) is an open source 3D creation software that provides native support for embedded Python scripting through its bpy application programming interface. In this work, all 3D structures are generated programmatically using Blender’s bpy package, enabling direct control over geometry construction, scene management, and rendering. This scripting interface allows generated structures to be validated and rendered automatically within a consistent environment. All original Blender scripts created for the bioinspired dataset are provided in the project GitHub. 

Rendering and validation are performed using a standardized Blender subprocess that clears the scene, instantiates generated objects, configures camera and lighting parameters, and produces rendered images. These canonical renders are used both for dataset validation and for downstream evaluation. The specific rendering and validation settings used throughout all Blender validation subprocesses is summarized in Table \ref{tab:bio3d-render}. To ensure consistency across samples, all renders use a fixed camera pose centered on the scene, standardized lighting, and automatically assigned grayscale materials. A script is considered valid if it executes successfully, produces at least one mesh object, and completes rendering without errors.

\begin{table}[h]
\centering
\small
\begin{tabular}{l c}
\toprule
\textbf{Render Property} & \textbf{Setting} \\
\midrule
Render engine & Eevee Next \\
Resolution & 1280 × 720 (PNG) \\
Camera pose & Fixed, TRACK\_TO scene center \\
Lighting & Single area light, standardized \\
Materials & Auto-assigned grayscale \\
\bottomrule
\end{tabular}
\caption{Rendering and validation settings used to produce canonical views for the Bio3D dataset.}
\label{tab:bio3d-render}
\end{table}

\subsection{Dataset Processing}
For the dataset processing pipeline, full notebooks with structured prompts are provided on the project GitHub. Additionally, the instruction/prompt generation files are also documented there.

\subsection{LLMs and VLMs}
BioinspiredLLM, first developed in \cite{Luu2023BioinspiredLLM:Materials}, here is based on the more recent Llama-3.1-8B-instruct architecture, specifically the model weights can be found on HuggingFace repository \texttt{lamm-mit/Llama3.1-8b-Instruct-CPT-SFT-DPO-09022024}. This specific version of BioinspiredLLM, was developed in \cite{Lu2024Fine-tuningCapabilities_updated} using continued pre-training, supervised fine-tuning, and direct preference optimization (DPO). The model used here in the Bioinspired123D pipeline is a quantized 4 bit GPT-Generated Unified Format (GGUF) version. The base model of Bioinspired3D is Llama-3.2-3B-instruct. Within the final agent system, the other agent models were GPT-4o-mini \cite{achiam2023gpt}.

\subsection{Model Training}
Bioinspired3D is developed by supervised finetuning the instruction tuned language model Llama-3.2-3B-Instruct on a curated dataset of Blender instructions. Each training example pairs a natural language prompt with a corresponding Blender Python script. Inputs are formatted using the Llama chat template (system, user, assistant), and training follows standard next token prediction, with loss masked over the prompt.

Finetuning and inference are implemented in Python using the Hugging Face Transformers library for model loading, tokenization, and training. Parameter efficient finetuning is performed with the PEFT library \url{(https://huggingface.co/docs/peft/en/index)} using LoRA adapters. Training is orchestrated with the Trainer API, with datasets managed using PyTorch and pandas. To reduce memory usage, mixed precision training (FP16) and gradient checkpointing are employed.

All finetuning experiments are conducted using PyTorch with GPU acceleration when available. Model weights and tokenizer configurations are sourced from the Hugging Face Hub, and all scripts and configuration files used for training and evaluation are provided in the project GitHub repository. 

\subsection{Retrieval-Augmented Generation (RAG)}
RAG was implemented for Bioinspired3D, and all agents including the  the Vision-Language Model agent. For Bioinspired3D and language model agents, a JSON file consisting of the base bioinspired scripts and respective captions was loaded. For the Vision-Language Model, a JSON consisting of bioinspired descriptive captions and file paths to a folder of generic renders of each bioinspired base model was similarly loaded. All RAG content was loaded using an embedding model (\url{BAAI/bge-small-en-v1.5}). All JSON files and renders are available on the project GitHub. 

\subsection{Benchmark}
A custom benchmark for bioinspired 3D structure generation was developed for this study. For consistency, all benchmark prompts follow the template ``Write a Blender script to make a (shape)'', where (shape) varies across a diverse set of tasks and difficulty levels. Although this template is used for evaluation, the model is finetuned to handle a broader range of natural language prompt formulations. The benchmark consists of 320 prompts spanning multiple categories. 
During evaluation, inference for Bioinspired3D is performed with a sampling temperature of 0.1. Model outputs are saved as text files, from which Blender Python scripts are extracted and written to .py files. These scripts are executed in a headless Blender subprocess using a standardized validation script. Validation success is assessed by parsing execution logs to determine whether a valid 3D structure is generated without runtime errors.

To evaluate structural correctness, rendered outputs are manually inspected and graded on a continuous scale from 0 to 1, where 0 indicates an incorrect or broken structure and 1 indicates a correct realization of the intended design. The full benchmark, including prompts and grading criteria, is provided in the project GitHub. While this benchmark relies on comparative human grading, future work could replace or augment human evaluation with vision–language models trained on graded renderings, or conditioned via retrieval on representative examples spanning the scoring range, enabling scalable evaluation while preserving alignment with human judgment.

\subsection{Agentic System}
The agentic system is implemented using LangGraph \url{(https://github.com/langchain-ai/langgraph)}, which formalizes multi-agent interaction as a graph structured process. In this framework, agents are represented as nodes, and directed edges define permissible transitions between agents based on the current design state. All agents operate over a shared design state that stores intermediate artifacts, including the design prompt, generated Blender script, render outputs, execution status, and evaluation feedback.

The system begins with Bioinspired3D, which generates an initial Blender Python script from the design prompt. The script is executed and rendered in a headless Blender subprocess, after which a vision language model evaluates the rendered output for physical validity and semantic alignment with the prompt. Based on this evaluation, the system transitions along one of several predefined paths. If the design is approved, the process terminates and the final render is returned. If execution errors occur, control is routed to the Repair agent that attempts to repair the script. If the render is valid but does not meet design intent, a Refinement agent modifies the script to better match the prompt.

This evaluation and refinement loop continues for a fixed number of iterations, after which the system terminates to prevent uncontrolled refinement. Throughout the process, all agents read and write to the same shared design state, ensuring consistent context and enabling coordinated decision making. The full implementation details, including agent logic and execution code, are provided in the project GitHub.

\subsection{3D Printing}
The bioinspired structures were generated in Blender via the Bioinspired3D and Bioinspired123D pipelines and exported as STL files for fabrication. Most black-colored samples were fabricated on a Stratasys F120 using ABS material with SR-30 dissolvable support. Remaining samples were printed on a Bambu X1C using either white PLA (Elegoo) or black PETG.

\section*{Acknowledgments}
The authors would like to thank Jesse de Alva for the help with 3D printing the bioinspired structures. This work was supported in part by Google, the MIT Generative AI Initiative, USDA (grant number 2021-69012-35978), with additional support from NIH. This material is based upon work supported by the National Science Foundation Graduate Research Fellowship under Grant number 2141064.

\section*{Author contributions}
R.K.L. developed the dataset, dataset processing pipeline, and custom benchmark, finetuned the models, developed the agentic system, and conducted assessments and analyses. R.K.L. also wrote the initial draft of the manuscript. M.J.B. supervised the project and contributed to its conceptual development. All authors contributed to editing and finalizing the manuscript.

\section*{Competing interests}
The authors declare no conflicts of interest.

\section*{Code and model weight availability}

All codes, protocols, notebooks, datasets, and Blender files developed in this study are available at:
\url{https://github.com/lamm-mit/Bioinspired123D}. Model weights can be accessed at: 
\url{https://huggingface.co/collections/lamm-mit/bioinspired123d-models-and-datasets}. 

\bibliographystyle{naturemag}
\bibliography{references-mendeley, references}  

\clearpage


\section*{Supplementary material}

# Bioinspired123D: Generative 3D Modeling System for Bioinspired Structures
# SUPPLEMENTARY INFORMATION [*]

**Rachel K. Luu**
Department of Materials Science and Engineering
Laboratory for Atomistic and Molecular Mechanics (LAMM)
Massachusetts Institute of Technology
Cambridge, MA, USA
https://orcid.org/0000-0002-7821-934X

**Markus J. Buehler**
Department of Civil and Environmental Engineering
Department of Mechanical Engineering
Center for Computational Science and Engineering
Schwarzman College of Computing
Laboratory for Atomistic and Molecular Mechanics (LAMM)
Massachusetts Institute of Technology
Cambridge, MA, USA
https://orcid.org/0000-0002-4173-9659

mbuehler@MIT.EDU

# 1 Supplementary Information

## 1.1 Model Training

Training dynamics, including loss, gradient norm, and total training steps, are shown in Figure 1a. To select the final checkpoint for the finetuned model, validation success rates across checkpoints are reported in Figure 1b. Here, validation success is defined solely as whether the generated Blender Python script executes without error and produces a runnable output, independent of whether the resulting geometry matches the intended design. To assess the impact of finetuning and retrieval, Blender execution success rates are compared across models in Figure 1c. While retrieval improves the base model's ability to generate executable scripts, the finetuned model consistently achieves higher execution success, demonstrating the combined benefits of supervised finetuning and retrieval augmentation.

## 1.2 Reasoning vs. Non-Reasoning

To assess the impact of explicit reasoning on code generation, we compare model outputs produced with and without a "Think step by step." prompt with examples shown below. Enabling explicit reasoning seems to primarily improves semantic alignment moreso than syntax validity as both generations execute correctly, but only the reasoning-enabled code produces geometry consistent with the prompt. In the non-reasoning setting, the generated script applies a sequence of boolean subtraction operations, but the resulting geometry remains visually indistinguishable from a solid sphere. This occurs because the cylindrical "tubules" are placed largely outside the effective interior of the sphere and are

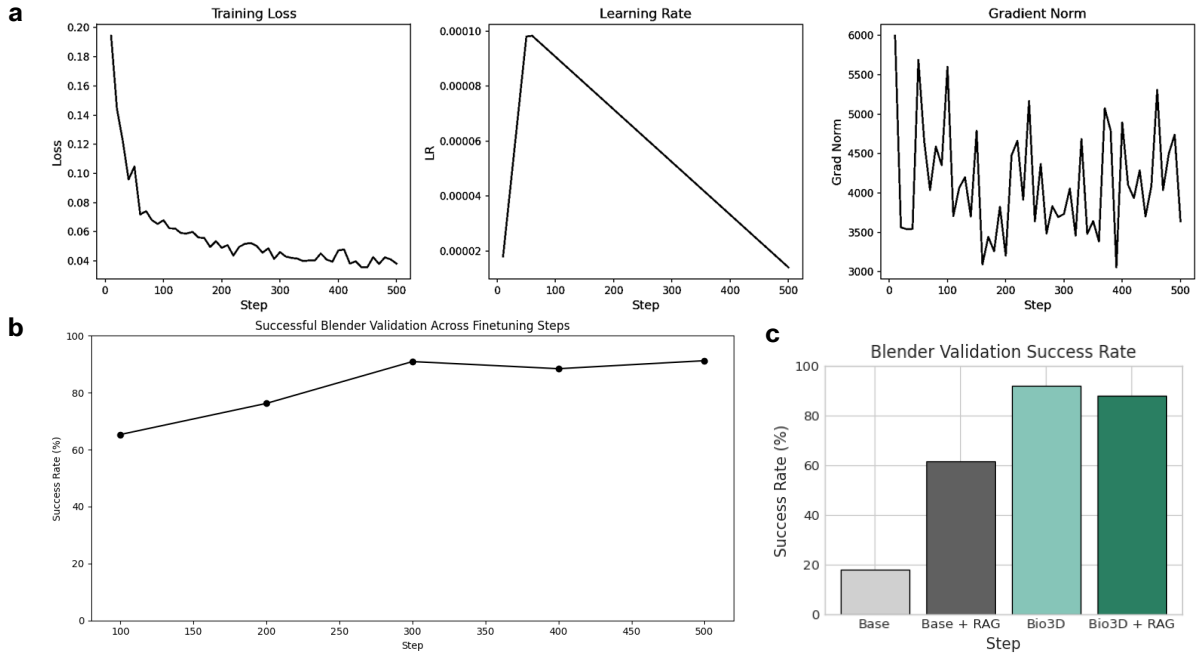

Figure 1: a) Training Loss, Learning Rate, and Gradient Norm across finetuning steps. b) Successful Blender Validations across finetuning steps where success rate(%) is measured as successfully validated scripts that ran properly in Blender and created an object. c) Bar plot comparison of Blender Validation success rate of the final finetuned checkpoint versus the base model with and without retrieval augmented generation (RAG) implementation, where RAG here is provided source of the base scripts.

applied sequentially without ensuring sufficient overlap or penetration depth, leading to negligible geometric removal after boolean application. As a result, the intended tubular structure is not realized in the final visual.

**Conversation with:** `lamm-mit/Bioinspired3D`

**User:** Write Blender code to make a tubular structure within a sphere instead of a cube.

**Bioinspired3D:**
```python
import bpy
import math

bpy.ops.object.select_all(action='SELECT')
bpy.ops.object.delete(use_global=False)

# Create a sphere
bpy.ops.mesh.primitive_uv_sphere_add(radius=2, location=(0, 0, 0))
sphere = bpy.context.object

# Define the number of tubules
count_tubules = 12

# Function to create tubules
def create_tubules():
    for i in range(count_tubules):
        angle = i * (2 * math.pi / count_tubules)
        x = math.cos(angle)
        y = math.sin(angle)
        bpy.ops.mesh.primitive_cylinder_add(radius=0.1, depth=3, location=(x, y, 0))
        tubule = bpy.context.object
        modifier = sphere.modifiers.new(name=f"Boolean_{i}", type='BOOLEAN')
        modifier.operation = 'DIFFERENCE'
        modifier.object = tubule
        bpy.context.view_layer.objects.active = sphere
        bpy.ops.object.modifier_apply(modifier=modifier.name)
        bpy.data.objects.remove(tubule, do_unlink=True)
```

```
# Create the tubules
create_tubules()
```

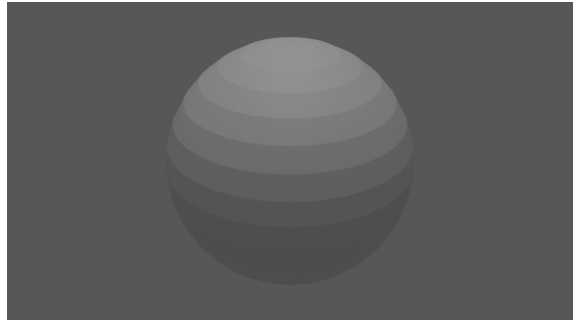

In contrast, the reasoning-enabled generation explicitly decomposes the task into intermediate design steps, including parameter definition, spatial layout of tubules, and controlled placement prior to boolean operations. This step-by-step structure leads to consistent intersection between the tubules and the base sphere, producing a clear internal tubular pattern that aligns with the prompt. These results suggest that explicit intermediate reasoning helps the model maintain geometric intent and avoid degenerate boolean operations.

**Conversation with:** `lamm-mit/Bioinspired3D`

User: Write Blender code to make a tubular structure within a sphere instead of a cube. Think step by step.

Bioinspired3D:
First, import the necessary libraries and clear the scene to start fresh.
```python
import bpy
import math

bpy.ops.object.select_all(action='SELECT')
bpy.ops.object.delete(use_global=False)
```

Next, define the parameters for the tubular structure. `num_tubules` controls how many tubules will be created, and `stretch_factor` determines how much the tubules will be stretched into ellipses.
```python
num_tubules = 12  # Total number of tubules
stretch_factor = 1.5  # Factor to stretch tubules into ellipses
```

Now, define a function that generates the tubules. This function will create a base sphere and then add tubules to it.
```python
def generate_tubules():
```

Inside the function, create a base sphere at the origin. This sphere will serve as the main structure from which the tubules will be subtracted.
```python
    bpy.ops.mesh.primitive_uv_sphere_add(radius=1, location=(0, 0, 0))
    sphere = bpy.context.object
```

Calculate the grid size for the tubules. The grid size is determined by taking the square root of the number of tubules and rounding up. Then, calculate the spacing between the tubules based on the grid size.
```python
    grid_size = math.ceil(math.sqrt(num_tubules))
    spacing = 2 / grid_size
    start_offset = -1 + spacing / 2
```

Generate the positions for each tubule in the grid. This creates a list of (x, y) coordinates for where each tubule will be placed.
```python
    positions = [(start_offset + (i % grid_size) * spacing,
                  start_offset + (i // grid_size) * spacing)
                 for i in range(num_tubules)]
```

```
Now, iterate over the generated positions to create and subtract each tubule from the sphere. For each position, add a
cylinder to represent a tubule.
```python
    for x, y in positions:
        bpy.ops.mesh.primitive_cylinder_add(radius=0.1, depth=3, location=(x, y, 0))
        tubule = bpy.context.object
```

Scale the tubule in the X-direction to stretch it into an ellipse based on the 'stretch_factor'.
```python
        tubule.scale[0] *= stretch_factor
```

Create a boolean modifier for the sphere to subtract the tubule. Set the operation to 'DIFFERENCE' and assign the
tubule object to the modifier.
```python
        bool_modifier = sphere.modifiers.new(name=f"Bool_{x}_{y}", type='BOOLEAN')
        bool_modifier.operation = 'DIFFERENCE'
        bool_modifier.object = tubule
```

Set the sphere as the active object and apply the boolean modifier. This will subtract the tubule from the sphere.
```python
        bpy.context.view_layer.objects.active = sphere
        bpy.ops.object.modifier_apply(modifier=bool_modifier.name)
```

Remove the tubule object from the scene after applying the modifier to keep the scene clean.
```python
        bpy.data.objects.remove(tubule, do_unlink=True)
```

Finally, call the function to generate the tubules.
```python
generate_tubules()
```
[...]

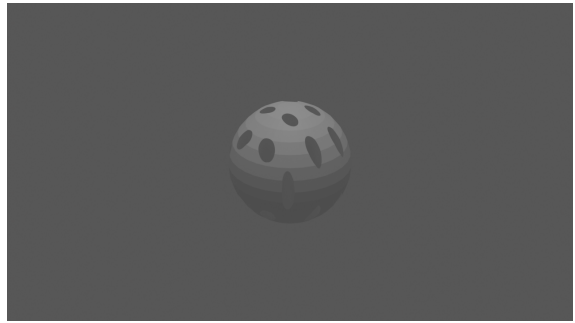

## 1.3 Retrieval Effects

To further examine how retrieval-augmented generation influences model behavior beyond aggregate metrics, we include qualitative examples here. Table 1 presents generations for a fixed prompt under varying retrieval depths k, alongside the specific retrieved scripts used as context. The prompt in this case asks the model to generate a helical structure with alternating rectangular and cylindrical fiber layers, which probes its ability to combine design elements across classes.

When no or limited retrieval is used (k=0, k=1), the model lacks access to concrete geometric priors and frequently produces syntactically invalid or incomplete Blender scripts. At k=2, the retrieved examples align closely with the prompt requirements: one script defines a helical structure composed of cylindrical fibers, while the other defines a helical structure composed of rectangular fibers. This pairing provides exactly the primitives needed for the task, allowing the model to successfully integrate both fiber types into alternating layers and produce a correct, executable result.

Increasing the retrieval depth beyond this point degrades performance. At k=3 and k=4, additional retrieved scripts introduce redundant or less relevant geometric patterns, which compete for attention and lead to structural ambiguity. In particular, at k=4, the model exhibits increased incoherence in both geometry composition and layer organization, despite having access to all necessary components. These examples indicate that retrieval effectiveness depends on script relevance, motivating the use of k=2 as a balanced operating point.

Prompt: `Write Blender code to make a helical structure made of both rectangular and cylindrical fibers, alternating layers.`

| k=0 (No RAG) Retrieved | k=2 Retrieved | k=2 Retrieved | k=3 Retrieved | k=4 Retrieved |
|---|---|---|---|---|
| None | Helical structure with cylindrical fibers | Helical structure with cylindrical fibers<br>Helical structure with rectangular fibers | Helical structure with cylindrical fibers<br>Helical structure with rectangular fibers<br>Helical structure | Helical structure with cylindrical fibers<br>Helical structure with rectangular fibers<br>Helical structure<br>Helical structure with noise |
| **Execution Error**<br>Script failed to run | **Execution Error**<br>Script failed to run | 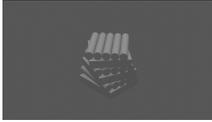 | 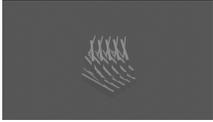 | 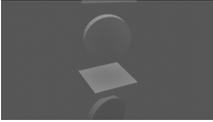 |
| **Status**<br>Fail (syntax/runtime) | **Status**<br>Fail (syntax/runtime) | **Status**<br>Pass (correct) | **Status**<br>Fail (incorrect) | **Status**<br>Fail (incorrect) |

Table 1: Qualitative examples illustrating the influence of retrieval-augmented generation on script execution and semantic correctness.

## 1.4 Quality vs. Compute

Figure 2 shows the relationship between average computational cost (as measured in tokens) and generation quality across Bioinspired3D, Bioinspired3D + RAG, and the full Bioinspired123D system. Computational cost is reported as the average total number of tokens processed per prompt (log scale), and performance corresponds to the benchmark.

Introducing retrieval augmentation leads to a moderate increase in quality relative to single-pass generation, with a corresponding increase in token usage. Extending the system to the full agentic Bioinspired123D pipeline further improves quality, though at substantially higher computational cost, mostly due to the VLM token cost. In this regime, token usage is dominated by iterative evaluation and refinement steps rather than initial generation.

Overall, this highlights a clear tradeoff between compute and performance, with higher-quality outputs associated with increased token budgets. This analysis provides context for the relative efficiency of different system components and motivates exploration of more compute-efficient refinement strategies.

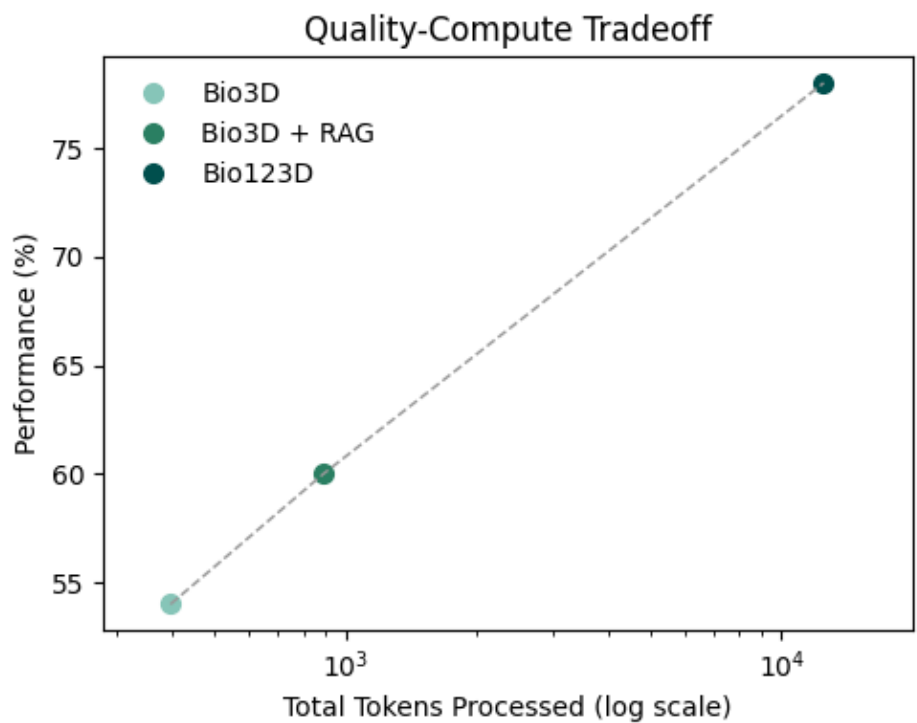

Figure 2: Quality–compute tradeoff across Bioinspired3D, Bioinspired3D + RAG, and Bioinspired123D. Points show mean quality scores as a function of average total tokens processed per prompt (log scale).


 
\end{document}